ARTICLE  OPEN 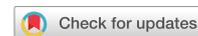

# Intrinsically weak magnetic anisotropy of cerium in potential hard-magnetic intermetallics

Anna Galler[1]✉, Semih Ener[2], Fernando Maccari[2], Imants Dirba[2], Konstantin P. Skokov[2], Oliver Gutfleisch[2], Silke Biermann[1,3,4,5] and Leonid V. Pourovskii[1,3]

Cerium-based intermetallics are currently attracting much interest as a possible alternative to existing high-performance magnets containing scarce heavy rare-earth elements. However, the intrinsic magnetic properties of Ce in these systems are poorly understood due to the difficulty of a quantitative description of the Kondo effect, a many-body phenomenon where conduction electrons screen out the Ce-4$f$ moment. Here, we show that the Ce-4$f$ shell in Ce–Fe intermetallics is partially Kondo screened. The Kondo scale is dramatically enhanced by nitrogen interstitials suppressing the Ce-4$f$ contribution to the magnetic anisotropy, in striking contrast to the effect of nitrogenation in isostructural intermetallics containing other rare-earth elements. We determine the full temperature dependence of the Ce-4$f$ single-ion anisotropy and show that even unscreened Ce-4$f$ moments contribute little to the room-temperature intrinsic magnetic hardness. Our study thus establishes fundamental constraints on the potential of cerium-based permanent magnet intermetallics.

npj Quantum Materials (2021)6:2 ; https://doi.org/10.1038/s41535-020-00301-6

## INTRODUCTION

The last decades have witnessed a rapidly growing demand for energy-efficient technologies, from optimized power generation and conversion devices to new transportation solutions. A key component of motors, generators, sensors, or actuators are high-performance permanent magnets. Identifying and designing improved materials for permanent magnet applications has thus the potential of huge economical and environmental savings[1,2]. Today, the world market for permanent magnets in the more advanced applications is largely dominated by rare-earth-based materials: the best hard-magnetic materials are based on the Nd-Fe-B "2–14–1" system whose extraordinary hard-magnetic properties are further boosted by partially substituting Nd with heavy rare-earth elements like Dy or Tb. In such rare-earth transition metal intermetallics, the electronic structure of the transition metal sublattice provides a large magnetization and high Curie temperature. The intrinsic magnetic hardness κ is due to the rare-earth local magnetic moment, which—due to the large spin–orbit coupling acting on the rare-earth electronic states—effectively converts the crystalline anisotropy into single-ion magnetic anisotropy[3].

Available resources of heavy rare earths are scarce, thus motivating efforts to replace heavy rare-earth intermetallics with so-called rare-earth balance magnets[2] containing more abundant rare-earth elements, in particular Ce[1,4–7]. A significant limitation to the magnetic hardness of Ce-transition metal intermetallics is caused by the well-known tendency of Ce-4$f$ electrons to form heavy-electron itinerant bands, as for instance observed in so-called "heavy-fermion" compounds[8]. Indeed, in the heavy-fermion regime the Ce local magnetic moment is screened by conduction electrons, which couple to the single Ce-4$f$ electron to form a nonmagnetic singlet state. Such a heavy-fermion behavior has been observed in various Ce intermetallics, for example, in the CeFe$_2$ Laves phase[9], as well as in the $R$Co$_5$ ($R$ = rare earth) and $R_2$Fe$_{14}$B families of hard-magnetic intermetallics[10,11] compromising, especially, the utility of Ce$_2$Fe$_{14}$B for magnetic applications[12].

In order to possess a significant single-ion magnetic anisotropy, an obvious conditio sine qua non is that the 4$f$ electron of Ce needs to be localized and its magnetic moment not significantly Kondo-screened. A reliable quantitative assessment of localized vs. heavy-fermion behavior of the Ce-4$f$ electrons is thus key to a successful theoretical search for prospective hard-magnetic cerium intermetallics. Such an assessment represents a formidable theoretical challenge. Some qualitative insights can be obtained from density functional theory (DFT) calculations treating Ce-4$f$ as itinerant bands[13], or estimating the evolution of the Kondo scale[14] from the DFT electronic structure. However, the formation of a heavy-fermion state in Ce-based compounds is a genuine non-perturbative many-body phenomenon, the quantitative description of which requires methods beyond DFT or DFT + U techniques[15,16]. At present, these limitations seriously hamper the possibilities of computational materials design in the field (see e.g., ref. [17]).

Here, we address this challenge by developing an ab initio approach to the magnetocrystalline anisotropy of Ce-based intermetallics. We describe the electronic structure of magnetic Ce-transition metal intermetallics, using advanced many-body electronic structure methods that take into account the effect of the Kondo screening. Furthermore, we supplement this first-principle framework with a variational approach to evaluate the magnitude of the Ce single-ion magnetic anisotropy explicitly taking into account the impact of the Kondo effect. In contrast to previous zero-temperature DFT calculations[17–19], the present study investigates the full temperature dependence of the Ce magnetic anisotropy. Our theoretical predictions are compared to the experimental magnetic anisotropy that we measure in the same wide temperature range.

[1]Centre de Physique Théorique, Ecole Polytechnique, CNRS, Institut Polytechnique de Paris, 91128 Palaiseau Cedex, France. [2]Functional Materials, Department of Material Science, Technische Universität Darmstadt, 64287 Darmstadt, Germany. [3]Collège de France, 11 place Marcelin Berthelot, 75005 Paris, France. [4]Department of Physics, Division of Mathematical Physics, Lund University, Professorsgatan 1, 22363 Lund, Sweden. [5]European Theoretical Spectroscopy Facility, 91128 Palaiseau, France. ✉email: galler.anna@gmail.com

We focus on the Ce–Fe "1–12" material system[4,7,17,20,21], which has recently been under scrutiny as a potential hard magnet[16,22,23]. Our calculations show a significant reduction of the Ce magnetic moment due to the Kondo effect. This Kondo screening is found to be very sensitive to substitutions and small-atom insertions. In particular, interstitial nitrogen is predicted to suppress the Ce contribution to the magnetization and magnetic anisotropy. This behavior is in drastic contrast to the one observed in "1–12" intermetallics with localized rare earths, like Nd, where nitrogenation has been found to significantly enhance the uniaxial magnetic anisotropy[24]. Indeed, the unexpected suppression of the Ce magnetization by N interstitials is absent if the Ce moment is treated as purely local and stems from a large enhancement of the Kondo scale by nitrogenation, as schematically shown in Fig. 1b–d. It is confirmed by our measurements of the anisotropy in polycrystalline samples with and without nitrogenation. Such interstitial atoms effectively controlling the state of the Ce-4$f$ shell could be used for fine-tuning and probing the complex many-electron physics of Ce. Most importantly, our calculations establish the salient fact that even in the absence of Kondo screening, the small size of the Ce-4$f$ spin moment necessarily leads to rather small values of the magnetic anisotropy at elevated temperatures. Our study, therefore, reveals significant limitations for the Ce contribution to the magnetic hardness at temperatures relevant for applications.

## RESULTS

### Kondo screening of the Ce magnetic moment in "1–12" systems

The "1–12" Ce–Fe intermetallics crystallize in the tetragonal lattice structure shown in Fig. 1a. A partial substitution of Fe with another transition metal (usually Ti) is necessary to stabilize this "1–12" phase in the bulk. Hence, we consider the realistic $CeFe_{11}Ti$ composition and, for the sake of comparison, also parent thermodynamically unstable $CeFe_{12}$. For the latter, we assume the same crystal structure. The magnetic properties of "1–12" rare-earth transition metal intermetallics can be modified by small-atom interstitials, in particular nitrogen. Nitrogenation has been found to significantly enhance the uniaxial magnetic anisotropy in $NdFe_{12}N$ and $NdFe_{11}TiN$ (ref. [24]). Nitrogenated Ce–Fe "1–12" compounds have been patented as prospective high-performance magnets[5], with N interstitials found to significantly increase their Curie temperature and magnetization[5,20,21]. Therefore, we also consider nitrogenated $CeFe_{11}Ti$ with N interstitials in the experimentally determined Wyckoff 2b position as shown in Fig. 1a.

The ordered Ce-4$f$ magnetic moment $M_z$ in $CeFe_{12}$, $CeFe_{11}Ti$, and $CeFe_{11}TiN$ as a function of temperature $T$ is displayed in Fig. 2a as symbols. This Ce moment is induced by the exchange field $B_{ex}$ generated by the ferromagnetically ordered iron sublattice, as schematically shown in Fig. 1b–d. In our ab initio approach, abbreviated below as DMFT(QMC), we take into account the magnetization of Fe within the local spin-density approximation (LSDA), while many-electron effects on the Ce-4$f$ shell are treated within dynamical mean-field theory (DMFT) in conjunction with a numerically exact quantum Monte-Carlo (QMC) approach[25]. With the Fe ferromagnetism included within LSDA, the exchange field $B_{ex}$ acting on the Ce-4f shell is virtually temperature independent (this approximation is justified since we focus on the temperature range up to 300 K, which is well below the Curie temperatures of 485 and 710 K for CeFe$_{11}$Ti and CeFe$_{11}$TiN, respectively). Hence, the $M_z$ vs. T curve reflects the temperature dependence of the Ce-4f magnetic susceptibility. One observes a weak T dependence for all three compounds; it is very different from a Curie-like behavior expected for Ce-4f local moments subjected to an exchange field. The obtained values are well below the saturated Russel-Saunders value $M_z = g_J \cdot J = 2.14 \mu_B$ for the Ce $f^{5/2}$ manifold. In particular, in CeFe$_{11}$TiN the 4f moment is tiny (0.04 $\mu_B$). A small increase of the Ce-4f magnetic moment with respect to pure CeFe$_{12}$ is induced by Ti substitution. Overall, the magnitude and temperature dependence of the Ce-4f magnetic moment in the investigated "1–12" intermetallics correspond to a Kondo-screened Fermi-liquid Ce-4f state.

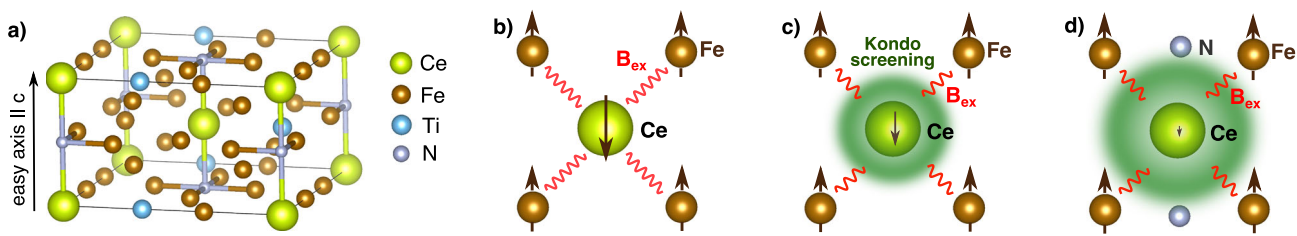

**Fig. 1 Crystal structure, Ce magnetism, and Kondo effect in Ce–Fe "1–12".** **a** Tetragonal body-centered crystal structure of $CeFe_{11}TiN$ (space group I4/$mmm$). **b** Quasi-atomic approximation to the Ce magnetism. The Ce-4$f$ spins are aligned antiparallel with respect to the Fe moments due to the exchange field $B_{ex}$. The magnitude of the Ce magnetic moment in this quasi-atomic picture is determined by $B_{ex}$ and the crystal field splitting on the Ce-4$f$ shell. **c** Many-electron effects drastically alter this picture: the Kondo screening (green cloud) reduces the Ce magnetic moment concomitantly lowering the Ce contribution to the magnetic anisotropy. **d** The Kondo effect is strongly sensitive to the local environment: it is dramatically enhanced by nitrogen interstitials thus almost completely suppressing the Ce-4$f$ contribution to the magnetic anisotropy.

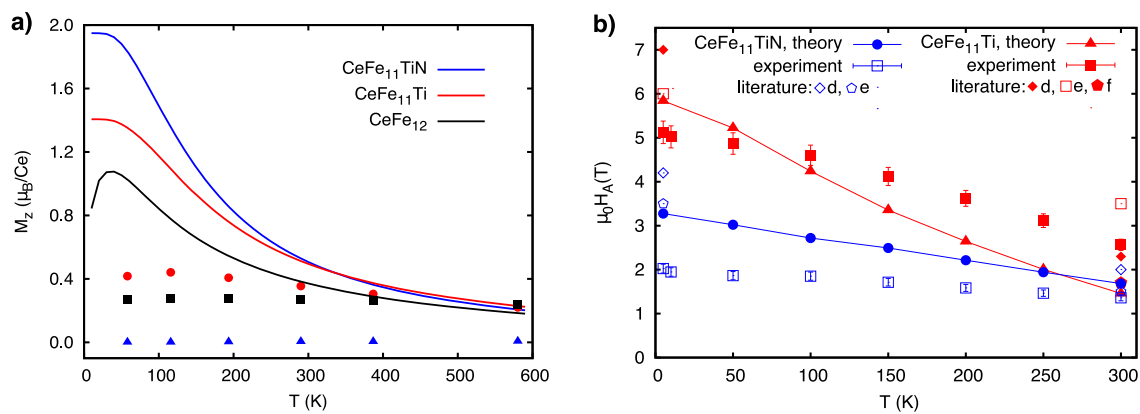

**Fig. 2 Temperature dependence of Ce-4$f$ magnetism in Ce–Fe "1–12".** **a** Calculated temperature dependence of the Ce-4$f$ magnetic moments $M_z$ in $CeFe_{12}$ (black squares), $CeFe_{11}Ti$ (red circles), and $CeFe_{11}TiN$ (blue triangles). At low temperatures the Kondo-screened Ce-4$f$ magnetic moments (symbols) are much smaller than their expected quasi-atomic values (solid lines). The Kondo effect is the strongest in $CeFe_{11}TiN$. **b** Temperature dependence of the anisotropy field $H_A$ (Tesla) in $CeFe_{11}Ti$ (in red) and $CeFe_{11}TiN$ (in blue). In addition to the theoretical and experimental results obtained in this work, values from (d) ref. [20], (e) ref. [21], and (f) ref. [4] are shown for comparison.

**Table 1.** Ce-4f magnetic moment $M_z$ ($\mu_B$) obtained in DMFT(QMC) (at 300 K), as well as the lowest-rank crystal field parameter $A_0^2\langle r^2\rangle$ (K), the exchange field $B_{ex}$ (Tesla), and the quasi-atomic Ce single-ion anisotropy constants $K_1^{at}$ [K] (at 4.2 and 300 K).

|  | $M_z$ | $A_2^0\langle r^2\rangle$ | $B_{ex}$ | $K_1^{at}$(4.2 K) | $K_1^{at}$(300 K) |
|---|---|---|---|---|---|
| CeFe$_{12}$ | 0.27 | −58 | 305 | −70.6 | −2.0 |
| CeFe$_{11}$Ti | 0.35 | 137 | 320 | 58.6 | 4.5 |
| CeFe$_{11}$TiN | 0.04 | 424 | 220 | 77 | 7.6 |

For comparison, the ordered Ce-4f magnetic moment $M_z^{at}$ has also been calculated within a quasi-atomic (Hubbard-I) approximation[26,27] to the Ce-4f quantum impurity problem (solid lines in Fig. 2a). In this approach, abbreviated below as DFT + HubI, the Kondo effect is neglected (as schematically shown in Fig. 1b). The Ce magnetic moment evaluated within DFT + HubI exhibits the expected Curie behavior and is significantly larger than the Kondo-screened one at low temperatures. Above $T \sim 400$ K, the magnitude of the Ce magnetic moments in CeFe$_{12}$ and CeFe$_{11}$Ti agree reasonably well between DFT + HubI and DMFT(QMC), indicating this temperature to be the Kondo scale of these compounds. Within the quasi-atomic picture one would expect the largest $M_z^{at}$ for CeFe$_{11}$TiN (blue solid line in Fig. 2a), for which the DMFT(QMC) moment remains, however, heavily screened in the whole temperature range up to 600 K.

Magnetic anisotropy of Ce in the quasi-atomic picture
In the quasi-atomic approximation (schematically shown in Fig. 1b), the single-ion anisotropy due to the Ce-4f magnetic moments is determined by $B_{ex}$ stemming from the Fe-sublattice and the crystal field (CF) splitting on the Ce-4f shell[28]. While the magnitude of $B_{ex}$ is similar in all three investigated compounds, the CF splitting turns out to be very different (see the crystal field parameters $A_0^2\langle r^2\rangle$ in Table 1). Of particular interest for the Ce-4f magnetic moment at low temperature is the composition of the ground state levels. The CF ground state in CeFe$_{11}$TiN is well separated from excited states and consists almost exclusively of the $|\pm 5/2\rangle$ eigenstates, which have the largest $|M_z^{at}|$ among the $j = 5/2$ manifold. CeFe$_{11}$TiN is thus expected to exhibit a strong easy-axis magnetic anisotropy in the quasi-atomic picture. In CeFe$_{11}$Ti, the contribution of the $|\pm 5/2\rangle$ eigenstates to the CF ground state is reduced leading to a smaller easy-axis anisotropy. In CeFe$_{12}$, the CF ground state is $|\pm 1/2\rangle$ corresponding to an easy-plane anisotropy (see Supplementary Fig. 1 for the full CF levels scheme). The peak of the CeFe$_{12}$ magnetic moment at ~40 K (Fig. 2a) can be explained by the fact that the applied exchange field oriented along the hard axis of CeFe$_{12}$ is not sufficient to saturate the ground state moment. Thus, with increasing temperature excited states with a larger moment along the hard axis become thermally populated before the usual Curie decay of the magnetization sets in.

We extract the values of the CF parameters, as well as $B_{ex}$ from our ab initio DFT + HubI calculations and then evaluate the Ce single-ion anisotropy constants. The resulting constants $K_1^{at}$ (in K per f.u.) are listed in Table 1. At liquid helium temperature, $K_1^{at}$ is easy-plane in CeFe$_{12}$, and easy-axis in CeFe$_{11}$Ti and CeFe$_{11}$TiN. For $T = 300$ K, CeFe$_{11}$TiN exhibits a substantially stronger easy-axis anisotropy as compared to CeFe$_{11}$Ti due to a larger value of its low-rank "20" crystal field parameter (see Supplementary Table I for values). Both compounds exhibit a strong reduction of the anisotropy with increasing temperature. The calculated room-temperature anisotropy constants $K_1^{at}$ are in very good agreement with the high-T limit analytical expression from ref. [28], which reads

$$K_1^{at} = -\frac{(J-1)(2J-1)(2J+3)}{20J} a_J A_2^0 \langle r^2 \rangle x^2, \quad (1)$$

with $x = \frac{2J(g_J-1)B_{ex}}{k_B T}$, where $g_J = 6/7$ is the gyromagnetic ratio for the Ce-4f$^1$ $J = 5/2$ shell. Using this analytical expression, one may understand the very rapid suppression of the Ce single-ion anisotropy apparent in Table 1. The Zeeman energy $2J|g_J - 1|B_{ex}$ associated to the exchange field $B_{ex} \sim 300$ T gives a characteristic temperature of $2J|g_J - 1|B_{ex}/k_B \sim 140$ K, explaining the rapid suppression of the $K_1^{at}$ values with temperature and the small value at room temperature. For comparison, we note that the same simple estimate for Sm$^{3+}$, with $g_J = 2/7$ gives a five times higher characteristic temperature (700 K), resulting in a 25 times slower reduction of $K_1^{at}$ vs. T at high temperatures. Hence, as the small magnitude of the spin represents an intrinsic feature of the Ce$^{3+}$ ion, i.e., not sensitive to the crystalline environment, the Ce-4f contribution to the magnetic hardness at room temperature and above is also expected to be intrinsically weak.

Figure 2b shows our measured anisotropy fields obtained from hard-axis measurements of the textured samples. Our results are consistent with the measurements at room temperature previously reported in refs. [4,20,21]. The anisotropy field is already relatively small at low temperature, and then exhibits a rather slow decay with temperature.

In order to compare theory and experiment, one needs to add the contribution from the Fe-sublattice to the calculated single-ion $K_1^{at}$ of Ce. As in this work, we do not aim at calculating the Fe-sublattice magnetic anisotropy, we extract its value and temperature dependence from measurements performed on YFe$_{11}$Ti (ref. [29]), using the experimental dependence $K_1^{Fe}(T) = 23.3 - 28T/T_c$ (K/f.u.) measured in ref. [29] together with our measured Curie temperatures. At low T, the iron contribution is much smaller than the Ce one and does not affect the picture qualitatively.

Using the relation $H_A \approx \frac{2K_1^{tot}}{M_S}$, where $H_A$ is the anisotropy field, $M_S$ the total magnetic moment of the compound saturated in the direction of $H_A$, and $K_1^{tot} = K_1^{Fe} + K_1^{at}$, we calculate the corresponding values of $H_A$ in the quasi-atomic approximation listed in Table 2 together with the experimental values. The quasi-atomic approximation predicts a strong anisotropy that decreases quickly with temperature, while experimentally one finds only a moderate decrease with T, in particular in CeFe$_{11}$TiN, which bears little resemblance to the predictions of DFT + HubI. Notice that the DFT calculations of ref. [18] treating Ce-4f as delocalized band-like states predict the zero-temperature $H_A$ of CeFe$_{11}$Ti to be reduced by nitrogenation by ~0.9 T, in disagreement with our measurements and previous experiments[20,21] finding the difference in $H_A$ between CeFe$_{11}$Ti and CeFe$_{11}$TiN$_x$ to be in the range of 2.5–3 T (see Table 2). The low-T anisotropy of CeFe$_{11}$Ti seems to be significantly underestimated by DFT[18,19]. Conversely, by treating Ce-4f as an open core Körner et al.[17] obtained $H_A = 18$ T for CeFe$_{11}$Ti, which is, similarly to our DFT + HubI result, strongly

| Table 2. Anisotropy field $H_A$ (Tesla). | | | | | | |
|---|---|---|---|---|---|---|
| T (K) | DFT | DFT + HubI | | QMC | | Experiment | |
| | 0 | 4.2 | 300 | 4.2 | 300 | 4.2 | 300 |
| CeFe$_{11}$Ti | 3.62$^a$ | 11.6 | 1.6 | 5.8 | 1.4 | 5.1 | 2.5 |
| | 2.83$^b$ | | | | | 7.0$^d$ | 2.3$^d$ |
| | 18$^c$ | | | | | 6.0$^e$ | 3.5$^e$, 1.7$^f$ |
| CeFe$_{11}$TiN | 2.77$^a$ | 12.3 | 2.5 | 3.2 | 1.6 | 2.0 | 1.3 |
| | | | | | | 4.2$^d$ | 2.0$^d$ |
| | | | | | | 3.5$^e$ | 1.5$^e$ |

Our values are obtained within the quasi-atomic approximation (DFT + HubI), as well as with the Kondo effect included using DMFT(QMC). Theoretical DFT values from refs. [18] ($^a$) and [19] ($^b$) are obtained with Ce-4f treated as valence states. DFT calculations of ref. [17] ($^c$) treated Ce-4f as open core. Experimental values are from this work, as well as from previous measurements of refs. [20], [21], and [4] that are indicated by superscripts d, e, and f, respectively. The experimental values in ref. [20] ($^d$) are for CeFe$_{11}$TiN$_{1.5}$. The low-temperature values in ref. [21] ($^e$) are for 1.5 K; we extracted their $H_A$ from the magnetization curves shown in Fig. 3 of this reference.

overestimated compared to experimental values. Hence, neither DFT nor the fully localized treatment of 4f is able to capture the impact of interstitials on the Ce magnetic anisotropy. As we show below, the theoretical picture is drastically modified when the impact of Kondo screening on the Ce-4f magnetic anisotropy is taken into account within a full many-body framework.

Ce magnetic anisotropy in the presence of Kondo screening
A quantitative evaluation of the magnetic anisotropy of a partially Kondo-screened magnetic moment is a formidable problem which—to the best of our knowledge—has not been addressed even for simple model systems so far. In the context of the present work on complex realistic systems, we have developed a variational approach to meet this challenge. In order to capture the interplay of the Kondo interaction ($H_K$) with the exchange ($H_B$) and crystal ($H_{CF}$) fields in the total impurity Hamiltonian $H = H_c + H_K + H_B + H_{CF}$ for the Ce-4f shell, we assume the ground state of the Ce-4f shell to be given by the following variational many-electron wave function:

$$\Psi = \sqrt{1-\alpha^2}\Psi_S + \alpha\Psi_J, \quad (2)$$

where $\Psi_S$ is the singlet formed by the Ce-4f total moment coupled to the conduction-electron sea defined by $H_c$; the second term $\Psi_J$ in Eq. (2) represents the 4f ground state in the absence of Kondo screening, i.e., determined by $H_B + H_{CF}$; $\alpha$ is a variational coefficient. Only the second term, due to its nonzero angular momentum, couples to the exchange field

$$\langle\Psi|H_B + H_{CF}|\Psi\rangle = \alpha^2\langle\Psi_J|H_B + H_{CF}|\Psi_J\rangle. \quad (3)$$

As shown in the Supplementary Note 3, the lowest-order anisotropy constant obtained with the variational ansatz Eq. (2) then reads

$$K_1 = K_1^{at}\frac{\alpha^2}{1+\delta}, \quad (4)$$

where $K_1^{at}$ is the anisotropy constant in the quasi-atomic approximation (DFT + HubI), as detailed above. The value of $\alpha^2$ is given in accordance with Eq. (3) by $M_z(T=0)/M_z^{at}(T=0)$ and can be extracted by extrapolating the curves in Fig. 2a to zero temperature. The denominator $1+\delta$ in Eq. (4), where $\delta = K_1^{at}/A$, is due to the change of $\alpha$ upon the rotation of $B_{ex}$ from the uniaxial $B_{ex}\|c$ to in-plane direction and can be evaluated from the dependence of $\alpha^2$ vs. $B_{ex}\|z$, calculated by DMFT(QMC) (see Supplementary Note 3 for details).

By multiplying the quasi-atomic $K_1^{at}$ for Ce with the corresponding reduction factors due to the Kondo screening (Eq. (4)) and adding the contribution of the Fe-sublattice $K_1^{Fe}(T)$, we obtain the total DMFT(QMC) anisotropy field listed in Table 2. Strictly speaking, the above variational treatment is applicable only at zero temperature. However, close to room temperature, the cerium contribution is overruled by the iron one, and the final value becomes relatively insensitive to the precise estimate of the Kondo reduction factor. At low temperature, instead, the Kondo reduction is very significant and the Ce contribution to $H_A$ is completely suppressed in the nitrogenated compound, but still relatively important in CeFe$_{11}$Ti. The values of the anisotropy field estimated within DMFT(QMC) are in good agreement with our experimental results and previous values reported in refs. [4,20,21], see Fig. 2b. In particular, the dramatic reduction of the low-T anisotropy field upon nitrogenation predicted by our calculations (by 45%, i.e., 2.6 T) is in quantitative agreement with the reduction observed in experiment. Our prediction for the low-T $H_A$ of CeFe$_{11}$Ti (5.8 T) is in the middle of the measured range (5.1–7.4 T), in contrast to the itinerant 4f-in-valence[18,19] and localized 4f-in-core[17] DFT limits, which are not able to capture the intermediate partially Kondo-screened state in this system. The itinerant picture with 4f-in-valence[18] performs, however, reasonably well for the strongly screened Ce-4f states in CeFe$_{11}$TiN.

Electronic structure and Ce-4f hybridization
As demonstrated above, the magnetic moment and anisotropy of Ce-4f in the CeFe$_{11}$MX intermetallics is heavily influenced by the Kondo effect. Here, we analyze its imprint on the electronic structure of CeFe$_{11}$Ti(N) and identify the origin of the enhanced Kondo screening in the nitrogenated system.

In Fig. 3a, we show the real part of the DMFT self-energy Re $\Sigma(\omega)$ in CeFe$_{11}$Ti and CeFe$_{11}$TiN, respectively, averaged over the six (degeneracy of the $J = 5/2$ multiplet) Ce-4f orbital contributions with the largest occupation number. The slope of Re$\Sigma(\omega)$ at $\omega = 0$ is a direct proxy for the strength of electronic correlations and determines the quasiparticle weight: $Z = \left[1 - \frac{d[\text{Re}\Sigma(\omega)]}{d\omega}|_{\omega=0}\right]^{-1}$. The obtained quasiparticle weight is higher in CeFe$_{11}$TiN, as compared to CeFe$_{11}$Ti, indicating a more delocalized character of the Ce-4f states in the nitrogenated compound. This can also be seen from the low occupation of $n = 0.86$ electrons on the Ce-4f shell in CeFe$_{11}$TiN, deviating stronger from the localized limit with $n = 1$ electron as compared to $n = 0.94$ in CeFe$_{11}$Ti. For a further comparison, in CeFe$_{12}$ the Ce-4f occupancy is $n = 0.93$, a bit lower than in CeFe$_{11}$Ti and thus indicating that the necessary Ti substitution slightly favors the localization of the Ce-4f states (for the spectral function of CeFe$_{12}$, we refer the reader to the Supplementary Fig. 4). The DMFT hybridization function shown in Fig. 3b quantifies the hybridization of the Ce-4f shell with other states of the system. Especially at low energies in the vicinity of the Fermi level, the hybridization function is much larger in CeFe$_{11}$TiN, indicating an increased mixing of the Ce-4f states with neighboring nitrogen states, e.g., N-2p, which is at the origin of the more delocalized behavior of the Ce-4f electrons.

The corresponding DMFT(QMC) spectral functions are displayed in Fig. 3c, d. In the characteristic three-peak structure of the Ce-4f spectral function the Kondo peak at the Fermi level is further split into two sub-peaks due to the spin–orbit splitting. The position of the localized features, i.e., the lower and upper Hubbard bands centered at −2 and 5 eV, respectively, is in good agreement with the photoemission (PES) spectra of CeFe$_2$ (ref. [9]; we are not aware of any experimental PES for the CeFe$_{12}$ materials family). The more delocalized character of the Ce-4f states in CeFe$_{11}$TiN is also reflected in the spectral function, Fig. 3d: a more prominent Kondo

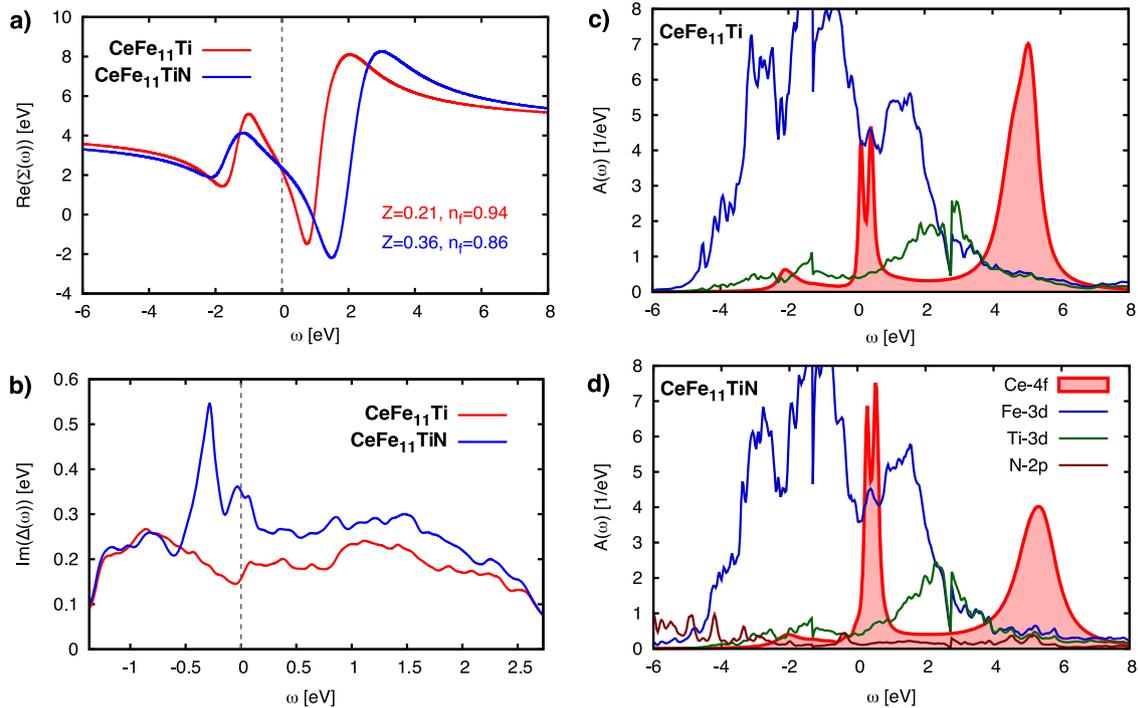

**Fig. 3 Impact of nitrogenation on electronic correlations and spectral function of CeFe$_{11}$Ti. a** Real part of the self-energy $\Sigma(\omega)$ in CeFe$_{11}$Ti (in red) and CeFe$_{11}$TiN (in blue). The quasiparticle weight $Z$ and Ce-4$f$ shell occupation $n_f$ are also indicated in the corresponding colors. **b** Imaginary part of the hybridization function $\Delta(\omega)$. In particular, at the Fermi energy Im$\Delta(\omega)$ is considerably larger in the nitrogenated compound. (The self-energies (**a**) and hybridization functions (**b**) are averaged over the six orbital contributions with the largest occupation number, respectively. For orbital-resolved self-energies, we refer the reader to the Supplementary Fig. 3.) **c, d** Corresponding spectral functions. The Ce-4$f$ Kondo peak at the Fermi energy is more pronounced in CeFe$_{11}$TiN due to the more delocalized character of its Ce-4$f$ electrons. For completeness, also the Fe-3$d$, Ti-3$d$, and N-2$p$ states are shown (the Fe-3$d$ contribution has been rescaled by a factor 1/3).

peak and less pronounced Hubbard bands signal a more delocalized behavior.

## DISCUSSION

This work introduces a DMFT-based approach to the Ce magnetization and magnetic anisotropy in ferromagnetic intermetallics. The intermediate valence behavior of Ce in the investigated intermetallics can be captured neither by pure DFT nor by approaches treating the Ce-4$f$ states as purely localized. Indeed, in our work, we show that a fully localized treatment of the Ce-4$f$ states within the quasi-atomic approximation strongly overestimates the Ce contribution to the anisotropy at low temperatures. The problem at hand thus calls for sophisticated many-body approaches, such as DMFT.

We derive a simple formula (Eq. (4)) relating the reduction of the magnetic anisotropy due to the Kondo screening to that of the ordered Ce moment; the latter being more easily accessible within DMFT. Moreover, this formula is of general applicability and can be used, for example, to extract the reduction of the anisotropy from the measured Ce magnetization. Our approach quantitatively captures the impact of Kondo screening on the intrinsic magnetic hardness $\kappa$ of realistic Ce systems, thus opening the possibility for a theoretical search of Ce-based intermetallics with enhanced $\kappa$.

Our calculations reveal a partially Kondo-screened state of the Ce-4$f$ magnetic moment in Ce–Fe intermetallics of the "1–12" family. Nitrogen interstitials strongly enhance the screening of the 4$f$ moment, thus completely suppressing the Ce single-ion contribution to the anisotropy both at liquid helium and room temperature. These theoretical predictions are validated by our experimental measurements finding the anisotropy field in CeFe$_{11}$Ti to be strongly suppressed by nitrogenation, in particular at low temperatures where the Ce contribution is the most significant. The effect of the dramatic Kondo-scale enhancement by nitrogen in the Ce–Fe "1–12" system indicates an unexpected sensitivity of this scale to interstitials in hard-magnetic intermetallics. The Kondo effect at low temperatures might thus be effectively controlled by such interstitials. Therefore, one may systematically search for interstitial elements that, in contrast to nitrogen in Ce–Fe "1–12", suppress the Kondo screening of the 4$f$ moment. The present theoretical framework can be employed in this search.

At the same time, the small magnetic anisotropy of a purely local Ce-4$f$ shell at elevated temperatures found in the present work limits the direct use of cerium-based "1–12" systems as hard-magnetic materials. Cerium can still be successfully used as a partial substitution element for the critical heavy rare-earth elements. In this case, the essential magnetic hardness is provided by heavy rare earths, but their concentration may in some cases be reduced without compromising the magnetic properties as recently shown in ref. [6] on the example of a mixture of La and Ce reducing the material cost of Nd$_2$Fe$_{14}$B-type hard-magnetic materials. However, the present study shows that cerium itself should not be expected to provide an essential contribution to the magnetic anisotropy at room temperature and above. Though in the present work we considered only the "1–12" family of Ce–Fe intermetalics, this conclusion based on the intrinsic properties of the Ce-4$f^1$ shell should be rather independent of the crystalline environment. It is expected thus to be of general validity for magnetic Ce intermetallics, unless one finds a—possibly exotic— mechanism allowing the hybridization to enhance the Ce-4$f$ anisotropy in the high-temperature local moment regime (i.e., above the Kondo temperature). Such an enhancement would manifest itself in a larger value of the Ce easy-axis magnetic moment obtained within a full DMFT treatment, as compared to the value obtained in the quasi-atomic approximation. In the current "1–12" systems, such an effect is not observed. Given the environmental and economical impact that is at stake if improved permanent magnet materials are found, it might however be worthwhile to systematically search for it.

## METHODS
### Theory

Our ab initio calculations—combined DFT and DMFT[27,30–32]—describe many-body effects in the Ce-4f shell taking into account on equal footing the key competing energy scales of the problem: the CF and exchange field induced by the spin-polarized electronic structure, as well as the spin–orbit coupling. We start from a charge-self-consistent spin-polarized DFT + DMFT calculation[33–37] of our target compounds, where we take into account the magnetization of Fe within the LSDA, with the Fe moment aligned along the $c||z$ axis, while we treat the Ce-4f magnetic moment within the quasi-atomic Hubbard-I approximation[26,27]. In our self-consistent DFT + DMFT calculations within the Hubbard-I approximation, abbreviated by DFT + HubI, we suppress the self-interaction contribution to the CF splitting, as well as the 4f contribution to the exchange field $B_{ex}$ on the Ce site. In result, the DFT exchange-correlation potential is calculated using a non-spin-polarized and spherical Ce-4f electronic density, while the nonspherical contributions and magnetization density of other states (including non-4f Ce orbitals like Ce-5d) are fully taken into account. This approach provides reliable CFs and $B_{ex}$ in hard-magnetic intermetallics, as previously demonstrated in ref. [16]. We employ the fully localized limit double-counting correction, in conjunction with the nominal $n = 1$ occupancy for the Ce-4f shell[38]. Since we obtain the Fe exchange field $B_{ex}$ within the DFT + HubI calculations, it is temperature independent. This is a reasonable approximation for the temperature range <300 K that we focus on. The experimental Curie temperatures $T_c$ for CeFe$_{11}$Ti and CeFe$_{11}$TiN of 485 and 710 K, respectively, are significantly >300 K. Our tests within the quasi-atomic treatment of the Ce-4f shell show that including the finite-temperature decay $B_{ex}$ leads to a rather small (<10% at 300 K) additional reduction of the Ce moment, as compared to the dependence (Fig. 2a) obtained with the fixed zero-T value of $B_{ex}$.

Having converged the self-consistent DFT + HubI calculations, we use the resulting charge density to produce the Kohn–Sham one-electron part of the Hamiltonian that includes the correct CF and $B_{ex}$ acting on the Ce-4f shell. We then perform an additional DMFT run using a hybridization-expansion continuous-time QMC solver in the segment picture[25,39]. We abbreviate this approach as DMFT(QMC). By using in DMFT(QMC) the charge and magnetization density obtained in DFT + HubI, we neglect the feedback of possible Ce-4f delocalization to the Fe magnetization and the exchange field $B_{ex}$. This treatment is in-line with the usual assumption that the 4f magnetism in the TM-RE hard magnets has a negligible influence on the TM-sublattice magnetism.

In the Ce-4f quantum impurity problem, we employ the basis of $f^1$ eigenstates of the atomic Hamiltonian, which represents two manifolds, $j = 5/2$ and $j = 7/2$, split by the spin–orbit coupling, with additional smaller splittings within the manifolds due to the CF and EF. We treat all 14 Ce-4f orbitals as correlated and employ the density–density approximation for the corresponding Coulomb vertex. We calculate the ordered Ce-4f magnetic moment $M_z = \text{Tr}\left[(\hat{L}_z + 2\hat{S}_z)\rho\right]$ from the converged DMFT density matrix $\rho$.

The Ce single-ion anisotropy constant within the atomic approximation, $K_1^{at}$, is evaluated from the dependence of the anisotropy energy $E_a^{at}$ on the azimutal angle $\theta$ of the exchange field $B_{ex}$ by fitting it to the form $E_a^{at} = K_1^{at} \sin^2\theta$. The dependence $E_a^{at}$ vs. $\theta$ is calculated by a numerical diagonalization of the Ce-4f CF Hamiltonian under $B_{ex}$ applied in the corresponding direction. Please refer to the Supplementary Note 2 for further details regarding our ab initio calculations and the calculation of the magnetic anisotropy.

### Experiment

Polycrystalline CeFe$_{11}$Ti samples are produced by the suction casting method using commercial purity elements (>99.99%). The as-cast samples are wrapped in Mo foil, sealed in a quartz ampule under argon atmosphere and annealed at 850 °C for 12 h followed by quenching in water. Part of the suction-cast plates are ground into powder <20 μm to carry out the nitrogenation. The determination of the crystal structures is done using a room-temperature x-ray powder diffractometer (XRD) and the XRD patterns are fitted with a ThMn$_{12}$-type (I4/mmm, space group no. 139) structure. Thermomagnetic measurements are carried out for the determination of the Curie temperatures. The anisotropy fields are measured on the polycrystalline CeFe$_{11}$Ti and CeFe$_{11}$TiN samples by using the method suggested by Durst and Kronmüller[40]. Applications of this model to ThMn$_{12}$ systems have been reported recently for similar material systems[41]. Further details of the sample production and characterization can be found in the Supplementary Note 1.

## DATA AVAILABILITY
The data generated and analyzed during this study are available from the corresponding author upon reasonable request.

## ACKNOWLEDGEMENTS
We thank Tilmann Hickel, Takashi Miyake, and Halil Sözen for fruitful discussions. This work was supported by the Deutsche Forschungsgemeinschaft and French Agence Nationale de la Recherche in the framework of the international collaborative DFG-ANR project RE-MAP (Project No. 316912154), the European Research Council under its Consolidator Grant scheme (Project No. 617196), and IDRIS/GENCI Orsay under Project No. t2020091393. A.G. acknowledges support through Schrödinger fellowship J-4267 of the Austrian Science Fund (FWF). L.V.P. acknowledges the support by the future pioneering program MagHEM, grant number JPNP14015, commissioned by the New Energy and Industrial Technology Development Organization (NEDO). We thank the computer team at CPHT for support.

## AUTHOR CONTRIBUTIONS
A.G. performed the electronic structure calculations, L.V.P. did the analytical derivations; S.E., F.M., I.D., and K.P.S. performed the experiments. L.V.P., S.B., and O.G. led the project. A.G., L.V.P., and S.E. drafted the manuscript. All authors discussed the results and contributed to the final version of the manuscript.

## COMPETING INTERESTS
The authors declare no competing interests.

## ADDITIONAL INFORMATION
**Supplementary information** is available for this paper at https://doi.org/10.1038/s41535-020-00301-6.

**Correspondence** and requests for materials should be addressed to A.G.

**Reprints and permission information** is available at http://www.nature.com/reprints

**Publisher's note** Springer Nature remains neutral with regard to jurisdictional claims in published maps and institutional affiliations.